\documentclass[pra,twocolumn,preprintnumbers,superscriptaddress]{revtex4}
\usepackage{amssymb}
\usepackage{amsmath}
\usepackage{amsfonts}
\usepackage{wasysym}
\usepackage{graphicx}
\usepackage{dcolumn}
\usepackage{bm}
\usepackage{color}
\usepackage[colorlinks,linkcolor=blue, urlcolor=blue, anchorcolor=blue, citecolor=blue]{hyperref}

\begin{document}

\title{Trapped Bose-Einstein Condensates with Attractive $s$-wave Interaction%
}
\date{\today }

\begin{abstract}
\end{abstract}

\begin{abstract}
Within the framework of the Gaussian-state theory, we show that the quantum many-body ground state of a trapped condensate with weakly attractive interaction is a single-mode squeezed vacuum state, as oppose to the coherent state under repulsive interaction. The spatial mode of the squeezed-state condensates satisfies a Gross-Pitaevskii like equation in which the interaction strength is augmented by a factor $3$ due to the large particle fluctuation of the squeezed state. We also study the collective excitations of the condensates by the tangential space projection, which leads to new two-particle excitations and confirms the phase transition from coherent-state to squeezed-state condensates. Our investigation clarifies the quantum states of the attractive condensates and will shed new light on research of the droplet phases in dipolar and multicomponent condensates.
\end{abstract}

\author{Tao Shi}
\email{tshi@itp.ac.cn}
\affiliation{CAS Key Laboratory of Theoretical Physics, Institute of Theoretical Physics,
Chinese Academy of Sciences, Beijing 100190, China}
\affiliation{CAS Center for Excellence in Topological Quantum Computation, University of
Chinese Academy of Sciences, Beijing 100049, China}
\author{Junqiao Pan}
\affiliation{CAS Key Laboratory of Theoretical Physics, Institute of Theoretical Physics,
Chinese Academy of Sciences, Beijing 100190, China}
\affiliation{School of Physical Sciences, University of Chinese Academy of Sciences,
Beijing 100049, China}
\author{Su Yi}
\email{syi@itp.ac.cn}
\affiliation{CAS Key Laboratory of Theoretical Physics, Institute of Theoretical Physics,
Chinese Academy of Sciences, Beijing 100190, China}
\affiliation{CAS Center for Excellence in Topological Quantum Computation, University of
Chinese Academy of Sciences, Beijing 100049, China}
\affiliation{School of Physical Sciences, University of Chinese Academy of Sciences,
Beijing 100049, China}
\maketitle

\textit{Introduction.}---The theoretical description of the quantum state of
Bose-Einstein condensates (BECs) has been of fundamental
importance \cite{BECbook1,Leggett,StringariRev}. Along with the development of
the superconducting theory \cite{BCS}, it was gradually realized that the
broken gauge symmetry (or the off-diagonal long-range order) and the phase coherence were the most essential ingredients of BECs~\cite{Penrose,Goldstone,ODLO,Anderson}, which allowed
the condensates to be described by macroscopic wavefunctions. The
experimentally demonstrated phase coherence of the BECs~\cite%
{coherence1,coherence2,coherence3} suggest that the coherent state might be
the most reliable representation for, in particular, open-system condensates~%
\cite{coherent_open}. In fact, the coherent-state description of BECs, including the Gross-Pitaevskii equation (GPE) and the Bogoliubov excitations~\cite%
{Bogoliubov} around the coherent-state condensates (CSCs), has achieved
great success in describing the trapped atomic BECs with repulsive
interactions. Although more sophisticated approaches~\cite{squpri95-3,Griffin,Stoof1,Stoof2,squaft95-5}
that incorporates corrections from Hartree-Fock-Bogoliubov terms were also
adopted to study the quantum states of BECs, both the conventional Bogoliubov treatment and these improved approaches are not fully self-consistent variational theory.

In this Letter, we revisit the quantum state of trapped BECs with contact interactions by employing the fully self-consistent Gaussian-state theory (GST)~\cite{Shi,Tommaso}, in which quantum many-body states are described within the whole Gaussian manifold. Surprisingly, we find that, as oppose to the CSCs under repulsive interactions, the quantum ground state of attractive BECs is a single-mode squeezed vacuum state. The spatial modes of the squeezed-state condensates (SSCs) satisfy a Gross-Pitaevskii like equation in which the interaction strength is augmented by a factor of $3$ compared to that of the GPE. The transition from CSCs to SSCs, realized by tuning the $s$-wave scattering length to a negative value, is of first order. Physically, SSCs are featured by their super-Poissonian particle-number statistics, in striking contrast to the Poissonian one for CSCs. Therefore, SSCs represent a new quantum state of macroscopic matter waves. We also study the collective excitations of the condensates via tangential space projection which takes into account both one-particle excitations (1PEs), i.e., conventional Bogoliubov excitations, and two-particle excitations (2PEs)~\cite{Tommaso}. Our approach naturally gives rise to the Goldstone zero mode in trapped Bose gases~\cite{Goldstone,Tommaso}. Moreover, in both CSC and SSC phases, we find low-lying 2PEs that are crucial for determining the properties of the condensates. We are aware of that number squeezing in the condensate mode was also considered by other researchers for repulsive interactions~\cite{squpri95-0,squpri95-1,squpri95-2,squpri95-3,squaft95-2,squaft95-4,squaft95-1,squaft95-3,squaft95-5,squaft95-6}, where the condensate should be dominated by coherent-state fraction.

\textit{Formulation.}---We consider a trapped condensate of $N$ interacting bosonic atoms at zero temperature. In second-quantized form, the Hamiltonian of the system is
\begin{align}
H=\int d{\mathbf r}\hat{\psi}^{\dagger }({\mathbf r}){\mathcal L}\hat{\psi}({\mathbf r})+\frac{U}{2}\int d{\mathbf r}\hat{\psi}^{\dagger 2}({\mathbf r})\hat{\psi}^{2}({\mathbf r}), 
\end{align}
where $\hat{\psi}({\mathbf r})$ is the field operator for bosonic atoms, ${\mathcal L}=-\hbar ^{2}\nabla ^{2}/(2m)+V({\mathbf{r}})-\mu$ is the single-particle Hamiltonian with $m$ being the mass of the atom, $V({\mathbf{r}})$ the external trap, and $\mu$ the chemical potential, and $U=4\pi\hbar ^{2}a_{s}/m$ represents the strength of the
collisional interaction with $a_{s}$ being the $s$-wave scattering length. Without loss of generality, we assume that the trapping potential is an isotropic harmonic oscillator, $V({\mathbf{r}})=m\omega_{\mathrm{ho}}^{2}{\mathbf r}^2/2$, where $\omega_{\rm ho}$ is the trapping frequency.

To proceed, let us briefly recall the GST~\cite{Shi,Tommaso}. A general Gaussian state takes the form%
\begin{equation}
\left\vert \Psi _{\mathrm{GS}}\right\rangle =e^{\hat{\Psi}^{\dagger }\Sigma
^{z}\Phi }e^{i\frac{1}{2}\hat{\Psi}^{\dagger }\xi \hat{\Psi}}\left\vert
0\right\rangle ,  \label{GS}
\end{equation}%
where $\hat{\Psi}({\mathbf r})=\big(\hat{\psi}({\mathbf r}),\hat{\psi}^{\dagger }({\mathbf r})\big)^{T}$ is the field operators in the Nambu basis and $\Sigma ^{z}=\sigma ^{z}\delta({\mathbf r}-{\mathbf r}^{\prime })$ with $\sigma ^{z}$ being the Pauli matrix.
In principle, the wave function $\Phi ({\mathbf r})=\langle \hat{\Psi}\rangle =\big(\phi({\mathbf r}),\phi^{*}({\mathbf r})\big)^{T}$ and the Hermitian matrix $\xi$ which
define the {\em Gaussian manifold} are the variational parameters to be
determined. Practically, instead of using $\xi$, we introduce the covariance
matrix $\Gamma ({\mathbf r},{\mathbf r}^{\prime })=\big\langle\{\delta \hat{\Psi}({\mathbf r}),\delta \hat{\Psi}^{\dagger }({\mathbf r}^{\prime })\}\big\rangle$\ of the fluctuation field $%
\delta \hat{\Psi}=\hat{\Psi}-\Phi $ to remove the gauge redundancy~\cite{Shi}, where $\Gamma$ and $\xi$ are related through the symplectic matrix $S\equiv e^{i\Sigma^z\xi}$ as $\Gamma=SS^\dagger$. For a Gaussian state, the coherent and squeezed parts of the condensates are characterized by $\Phi $ and $\Gamma$, respectively. It should be noted that, for short-hand notation, the products in Eq.~\eqref{GS} should be understood as the matrix multiplications in the coordinate and Nambu spaces~\cite{SM}.

With respect to the Gaussian state (\ref{GS}), Wick's theorem leads to the
mean-field Hamiltonian~\cite{SM},
\begin{equation}
H_{\mathrm{MF}}=E+(\delta \hat{\psi}^{\dagger }\eta +\eta ^{\ast }\delta
\hat{\psi})+\frac{1}{2}:\delta \hat{\Psi}^{\dagger }\mathcal{H}\delta \hat{%
\Psi}:,  \label{hmeanf}
\end{equation}%
where $E=\langle \Psi _{\mathrm{GS}}|H|\Psi _{\mathrm{GS}}\rangle $ is the
variational energy,
\begin{align}
\eta \lbrack \phi ,\Gamma ]& \equiv \left[ {\mathcal L}+U\left\vert \phi
({\mathbf r})\right\vert^{2}+2UG({\mathbf r},{\mathbf r})\right] \phi({\mathbf r})  \notag \\
& \quad +UF({\mathbf r},{\mathbf r})\phi ^{\ast}({\mathbf r}),  \label{yita}
\end{align}%
is the driving vector, and $\mathcal{H}[\phi ,\Gamma ]\equiv
\begin{pmatrix}
\mathcal{E} & \Delta \\
\Delta ^{\dagger } & \mathcal{E}^{\ast }%
\end{pmatrix}$ is a matrix with elements $\mathcal{E}[\phi ,\Gamma ]\equiv
{\mathcal L}+2U\big[|\phi({\mathbf r})|^{2}+G({\mathbf r},{\mathbf r})\big]$ and $\Delta \lbrack \phi
,\Gamma ]\equiv U\left[\phi^{2}({\mathbf r})+F({\mathbf r},{\mathbf r})\right] $. Here, $%
G({\mathbf r},{\mathbf r}^{\prime })\equiv \left\langle \delta \hat{\psi}^{\dagger }({\mathbf r}^{\prime
})\delta \hat{\psi}({\mathbf r})\right\rangle $ and $F({\mathbf r},{\mathbf r}^{\prime })\equiv
\left\langle \delta \hat{\psi}({\mathbf r}^{\prime })\delta \hat{\psi}({\mathbf r})\right\rangle
$ are the normal and anomalous Green functions, respectively. It should be noted that, in Eq.~\eqref{hmeanf}, the normal-ordered operators $:\hat O :$ is defined with respect to the Gaussian state.

The ground-state solution, $(\phi _{0},\Gamma _{0})$, can be obtained by numerically evolving the imaginary-time equations of motion (EOM)~\cite{Shi,Tommaso},
\begin{subequations}
\label{Imv}
\begin{align}
\partial _{\tau }\Phi & =-\Gamma
\begin{pmatrix}
\eta \\
\eta ^{\ast }%
\end{pmatrix}%
,  \label{Imva} \\
\partial _{\tau }\Gamma & =\Sigma ^{z}\mathcal{H}\Sigma ^{z}-\Gamma \mathcal{%
H}\Gamma ,  \label{Imvb}
\end{align}
\end{subequations}
in, for instance, a truncated harmonic oscillator basis~\cite{SM}, which converge at large imaginary time $\tau$. Equivalently, we may also find $(\phi _{0},\Gamma _{0})$ via diagonalizing the mean-field Hamiltonian $H_{\mathrm{MF}}$ by requiring $\eta \lbrack \phi _{0},\Gamma _{0}]=0$ and $S_{0}^{\dagger }\mathcal{H}[\phi _{0},\Gamma _{0}]S_{0}=I_{2}\otimes D$,
where $I_{2}$ is the $2\times 2$ identity matrix, $D$ is a diagonal matrix,
and $\Gamma _{0}=S_{0}S_{0}^{\dagger }$ is constructed self-consistently by
the symplectic matrix $S_{0}$ satisfying $S_{0}\Sigma ^{z}S_{0}^{\dag
}=\Sigma ^{z}$~\cite{Shi,Tommaso}.

\begin{figure}[ptb]
\centering
\includegraphics[width=0.9\columnwidth]{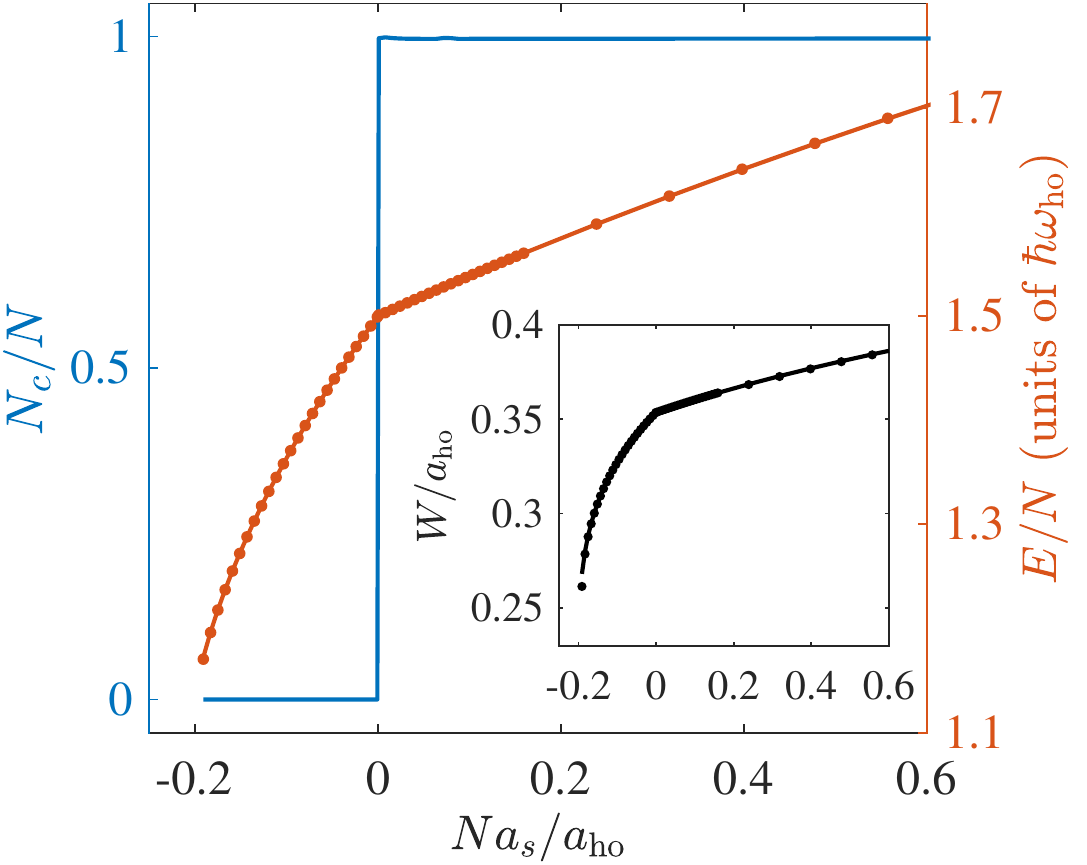}
\caption{(color online). $N_c/N$ (blue, left $y$ axis) and $E/N$ (red, right
$y$ axis) versus $a_s$ for 3D (a) and quasi-2D (b) traps. The insets show $W$
as a function of $a_s$. Solid lines are obtained via GST; while dots are
found by solving GPE for $a_s>0$ and Eq.~\eqref{Ims} for $a_s<0$.}\label{ground}
\end{figure}

\textit{CSC-SSC transition.}---Here we explore the ground-state properties of a weakly interacting condensate. Figure~\ref{ground} shows the coherent-state fraction, $N_{c}/N$,
and the energy per particle, $E/N$, versus the dimensionless interaction strength, $Na_{s}/a_{\rm ho}$, where $N_{c}=\int
d{\mathbf r}\left\vert \phi ({\mathbf r})\right\vert ^{2}$ is the particle number in coherent state and $a_{\rm ho}=\sqrt{\hbar/(m\omega_{\rm ho})}$ is the harmonic oscillator length. As can be seen, $N_{c}/N$ drops abruptly at $a_{s}=0$ from
essentially unit to zero as $a_{s}$ is tuned from positive to negative,
indicating that the variational ground state changes from a coherent state to a
squeezed vacuum state. This transition is further confirmed, in Fig.~\ref%
{ground}, by the non-analytic behavior of $E/N$ at $a_{s}=0$. Furthermore,
it is found that $E/N$ diverges if $Na_{s}/a_{\mathrm{ho}}\apprle-0.19$, signaling the collapse of system.

For the coherent-state solutions covered by our numerical calculations, the depletion is
always negligible. Thus, to the lowest order, the ground-state can be
expressed as $\left\vert \Psi _{\mathrm{CS}}\right\rangle =\exp\big[\int
d{\mathbf r}\phi _{0}({\mathbf r})\hat \psi^{\dagger }({\mathbf r})\big]\left\vert 0\right\rangle $, where $%
\phi _{0}$ is the solution of the equation $\eta =0$ with $G({\mathbf r},{\mathbf r})$ and $F({\mathbf r},{\mathbf r})$ being ignored, i.e., the conventional GPE. Interestingly, the corrections to $\phi_0({\mathbf r})$ can be systematically included by iteratively solving Eqs.~\eqref{Imv}. For demonstration purposes, here we show how to obtain the first-order correction under local density approximation (see Supplemental Material~\cite{SM} for details). To this end, we first diagonalize $\mathcal{H}[\phi_{0}({\mathbf r}),\Gamma=I_{2}\delta ({\mathbf r}-{\mathbf r}^{\prime })]$ with the trapping potential being ignored through Bogoliubov transformation. The resulting one-particle excitation (1PE) spectrum then give rise to $G({\mathbf r},{\mathbf r})=8[n_{0}({\mathbf r})a_{s}]^{3/2}/(3\sqrt{\pi })$ and $F({\mathbf r},{\mathbf r})=8[n_{0}({\mathbf r})a_{s}]^{3/2}/\sqrt{\pi }$ under local density approximation, where $n_{0}({\mathbf r})=|\phi _{0}({\mathbf r})|^{2}$. Substituting $G({\mathbf r},{\mathbf r})$ and $F({\mathbf r},{\mathbf r})$ back into the equation $\eta =0$, one obtains the GPE with Lee-Huang-Yang (LHY) correction~\cite{LHY,dipolarTh,twocomponentTh}, an equation that yields the first-order correction of $\phi_0$. From the above analysis, it is clear that the LHY-corrected GPE is applicable only when the condensate is dominated by the coherent-state fraction.

For the squeezed-state phase, our numerical results unveil that the correlation functions can always be diagonalized into $G({\mathbf r},{\mathbf r}^{\prime})\approx Nf({\mathbf r})f({\mathbf r}^{\prime})$ and $F({\mathbf r},{\mathbf r}^{\prime})\approx \sqrt{N(N+1)}f({\mathbf r})f({\mathbf r}^{\prime})$ with $f({\mathbf r})$ being the mode function. This remarkable result implies that the quantum many-body ground state of a SSC is in a single-mode squeezed vacuum state,
\begin{align}
\left\vert \Psi_{\mathrm{SS}}\right\rangle =\exp\left[\frac{1}{2}\xi_{0}(\hat b^{\dagger 2}-\hat b^{2})\right]\left\vert 0\right\rangle,
\end{align}
where $\hat b^{\dagger}=\int d{\mathbf r}f({\mathbf r})\hat{\psi}^{\dagger}({\mathbf r})$ and $\sinh \xi _{0}=\sqrt{N}$.

To gain more insight into the squeezed state, we compute the mean-field energy $E$ using wave
functions $\left\vert \Psi _{\mathrm{CS}}\right\rangle $ and $\left\vert
\Psi _{\mathrm{SS}}\right\rangle $, separately. It turns out that the ground-state energies take a unified form
\begin{equation}
E[\bar{\phi}({\mathbf r})]=\int d{\mathbf r}\bar{\phi}^{\ast}({\mathbf r})\left[ {\mathcal L}+\frac{U_{%
\mathrm{eff}}}{2}\left\vert \bar{\phi}({\mathbf r})\right\vert ^{2}\right] \bar{\phi}({\mathbf r}),  \label{eunfied}
\end{equation}%
where $\bar\phi=\bar{\phi}_{\mathrm{CS}}\equiv\phi _{0}$ and $U_{\mathrm{eff}%
}=U$ for coherent states; while $\bar\phi=\bar{\phi}_{\mathrm{SS}}\equiv\sqrt{N}%
f$ and $U_{\mathrm{eff}}=3U[1+(3N)^{-1}]\approx 3U$ for
single-mode squeezed states. We remark that the factor of $3$ enhancement of
$U_{\mathrm{eff}}$ for the SSC is contributed by the
Hatree-Fock-Bogoliubov terms, which, as shall be shown, also originates
from the large particle number fluctuation in squeezed states. A immediate
consequence of Eq.~\eqref{eunfied} is that the coherent (squeezed) state has
a lower energy when $a_{s}>0$ ($a_{s}<0$), in consistency with the
first order phase transition at $a_{s}=0$. More remarkably, the variational
principle, $\delta E[\bar{\phi}_{\mathrm{SS}}]/\delta \bar{\phi}_{\mathrm{SS}%
}^{\ast }=0$, leads to an effective equation%
\begin{equation}
\left[{\mathcal L}+3U\left\vert \bar{\phi}_{\mathrm{SS}}({\mathbf r})\right\vert ^{2}%
\right] \bar{\phi}_{\mathrm{SS}}({\mathbf r})=0,  \label{Ims}
\end{equation}%
for the wave function of SSCs. Apparently, Eq.~\eqref{Ims} has the same form as
the GPE that describes the CSCs except for that the interaction strength is
now tripled.

In Fig.~\ref{ground}, we compare the ground-state energy, $E/N$, and the
condensate width, $W=[\int d{\mathbf r}{\mathbf r}^{2}|\bar{\phi}({\mathbf r})|^{2}]^{1/2}$,
separately computed via the GST and the effective equations, i.e., GPE for $%
a_{s}>0$ and Eq.~\eqref{Ims} for $a_{s}<0$. As can be seen, two approaches
agree with each other for small $|a_{s}|$. However, visible discrepancy is
found close to the stability boundary. This discrepancy can be attributed to
the insufficient basis states used in GST calculations, which, in below,
will be further explored by examining the excitation spectrum of the system.
More interestingly, the similarity between Eq.~\eqref{Ims} and GPE allows us
to make an inference on the stability of a SSC based on that of a CSC~\cite{Ruprecht,Wieman}: for negative scattering length, the condensate remains metastable for $N|a_{s}|/a_{\mathrm{ho}}\apprle0.19$. Consequently, SSCs can only sustain finite number of particles for a given negative scattering length $a_s$. 

\textit{Density fluctuations}.---We now explore the collective excitations around a steady-state
solution $(\phi _{0},\Gamma _{0})$ by the tangential space projection approach~\cite{Tommaso}. For this purpose, we linearize the real-time EOM~\cite{Shi}
\begin{subequations}
\label{Rev}
\begin{align}
i\partial _{t}\phi & =\eta ,  \label{Reva} \\
i\partial _{t}\Gamma & =\Sigma ^{z}\mathcal{H}\Gamma -\Gamma \mathcal{H}%
\Sigma ^{z},  \label{Revb}
\end{align}%
around $(\phi _{0},\Gamma _{0})$ by letting $\phi=\phi_0+\delta\phi$ and $\Gamma=\Gamma_0+\delta\Gamma$, where $\delta \phi $ represents the 1PEs and $\delta \Gamma $ relates to the 2PEs, $\delta \xi$ (the fluctuation of $\xi$), according to~\cite{SM}
\end{subequations}
\begin{equation}
\delta \Gamma =2%
\begin{pmatrix}
\delta G({\mathbf r},{\mathbf r}^{\prime }) & \delta F({\mathbf r},{\mathbf r}^{\prime }) \\
\delta F^{\dagger }({\mathbf r},{\mathbf r}^{\prime }) & \delta G({\mathbf r}^{\prime },{\mathbf r})%
\end{pmatrix}%
=2iS_{0}%
\begin{pmatrix}
0 & \delta \xi \\
-\delta \xi ^{\dagger } & 0%
\end{pmatrix}%
S_{0}^{\dagger }.  \notag
\end{equation}%
It then follows from the linearization of Eqs.~\eqref{Rev} that the
fluctuations obey the equations
\begin{subequations}
\label{fRev}
\begin{align}
i\partial _{t}\delta \phi & =\mathcal{E}\delta \phi +\Delta \delta \phi
^{\ast }+2U\phi _{0}\delta G({\mathbf r},{\mathbf r})  \notag \\
& \quad +U\phi _{0}^{\ast }\delta F({\mathbf r},{\mathbf r}),  \label{fReva} \\
i\partial _{t}\delta \xi & =\{D,\delta \xi \}-i(S_{0}^{\dagger }\delta
\mathcal{H}S_{0})_{12},  \label{fRevb}
\end{align}%
where the subscript `$12$' denotes the off-diagonal block in the Nambu basis
and $\delta \mathcal{H}=%
\begin{pmatrix}
\delta \mathcal{E} & \delta \Delta \\
\delta \Delta ^{\dagger } & \delta \mathcal{E}%
\end{pmatrix}%
$ with $\delta \mathcal{E}=2U[\phi _{0}^{\ast }({\mathbf r})\delta \phi({\mathbf r})+\phi
_{0}({\mathbf r})\delta \phi ^{\ast }({\mathbf r})+\delta G({\mathbf r},{\mathbf r})]$ and $\delta \Delta
=U[2\phi _{0}({\mathbf r})\delta \phi ({\mathbf r})+\delta F({\mathbf r},{\mathbf r})]$. Equations~\eqref{fRev}
and their conjugate counterparts constitute the generalized Bogoliubov
fluctuation analysis~\cite{Tommaso}, which is equivalent to the random-phase
approximation extensively used in condensed matter physics~\cite%
{CRPA,Demler,Tommaso}. Numerically, we diagonalize the linearized Eqs.~%
\eqref{fRev} in a truncated harmonic oscillator basis to find the excitation spectrum~\cite{SM}. It should be noted that, owing to the rotational symmetry, each excitation can be labeled by its orbital angular momentum $L$.

\begin{figure}[ptb]
\centering
\includegraphics[width=0.75\columnwidth]{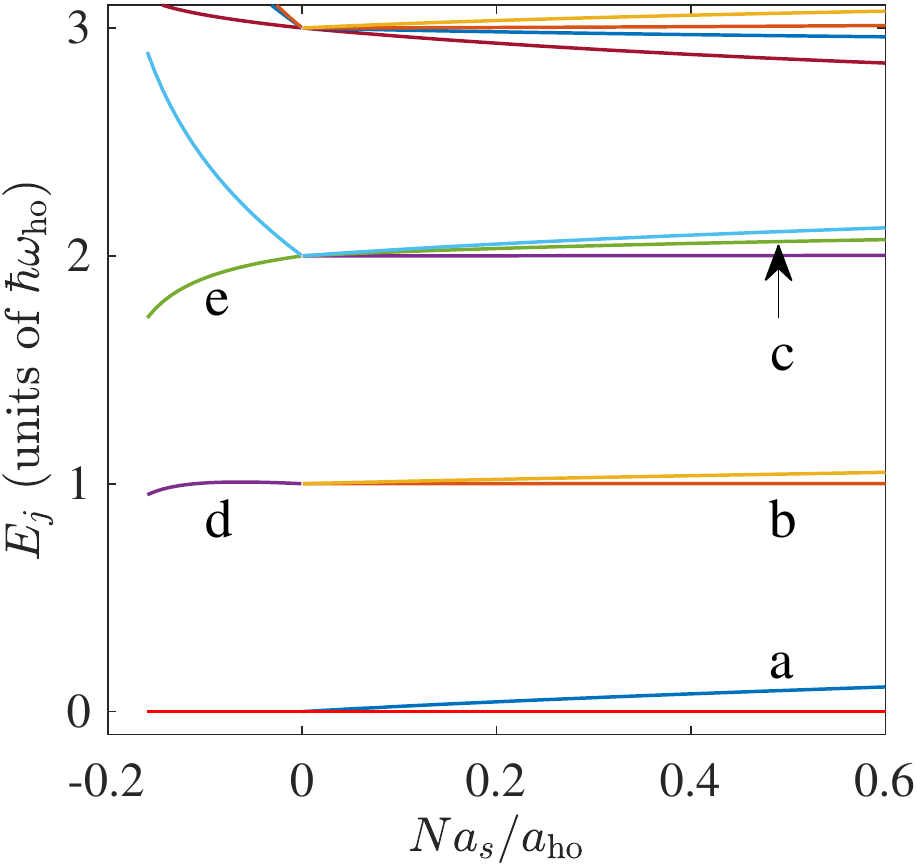}
\caption{(color online). Energies of the low-lying density excitations across the CSC-SSC
transition..}\label{spectra}
\end{figure}

Figure~\ref{spectra} shows the low-lying energy spectrum across the CSC-SSC transition for the excitations with total angular momentum $L=0$ and $1$. A immediate observation is that, independent of $a_s$, a Goldstone zero mode associated with the U(1) symmetry breaking always presents. This result is highly nontrivial as in the general situation, $\Gamma _{0}\neq I_{2}\delta ({\mathbf{r}}-{\mathbf{r}}^{\prime })$, the mean-field Hamiltonian $\mathcal{H}[\phi _{0},\Gamma _{0}]$ is gapped, which  seems violating the Goldstone theorem. Previously, this inconsistency was usually remedied by introducing approximations to modify $\mathcal{H}$~\cite{squpri95-1,squpri95-2,squpri95-3,squaft95-5,squaft95-6,Stoof1,Stoof2} such that the Hugenholtz-Pines condition~\cite{HP1,HP2} is satisfied. Here, instead of diagonalizing the mean-field Hamiltonian, our approach diagonalizes the linearized EOM, Eqs.~\eqref{fRev}, which computes the excitation self-consistently by taking a full consideration of 1PEs, 2PEs, and their couplings. As analyze by Guaita {\it et al}.~\cite{Tommaso} in lattice systems and also confirmed by our numerical computation for trapped BECs, the tangential space projection approach gives rise to the Goldstone zero mode natually.

We now turn to study the nonzero modes of the CSC phase. Here the density fluctuation, $\delta n=\phi _{0}\delta \phi ^{\ast }+\phi _{0}^{\ast }\delta \phi +\delta G({\mathbf r},{\mathbf r})$, is dominated by the 1PEs which couple to the 2PEs as described by Eqs.~\eqref{fRev}. The lowest nonzero mode (labeled a in Fig.~\ref{spectra}) is a two-particle breathing mode which represents the radial expansion and contraction of the gas, in analogy to the single-particle Bogoliubov breathing mode. As $a_{s}$ is lowered, mode a softens toward zero energy, signaling the onset of the instability of the CSC phase. The excitations around $\hbar\omega_{\rm ho}$ contains three 1PEs and three 2PEs which are six-fold degenerate at $a_{s}=0$ owing to the rotational symmetry of the system. After interaction is switched on, the degeneracy of these modes is partially lifted by the 1PE-2PE coupling. As a result, these six modes are evenly grouped into two branches, each of which is of three-fold degenerate. Particularly, excitations in the lower branch (labeled b) are single-particle dominating dipole modes, representing the center-of-mass motion of the condensate in an isotropic harmonic potential. As expected, the excitation energy of these dipole modes, $\hbar \omega _{\mathrm{ho}}$, is independent of the scattering length. The excitations around $2\hbar \omega _{\mathrm{ho}}$ are grouped into three branches: the middle branch (labeled c) which is nondegenerate and is dominated by the 1PE can be identified as the the conventional breathing mode; while other two branches that are of three-fold and five-fold degeneracy are dominated by the 2PEs. 

In the SSC phase, the vanishing $\phi _{0}$ has two implications: i) 1PEs and 2PEs are decoupled and ii) the density fluctuation $\delta n=\delta G({\mathbf r},{\mathbf r})$ only consists of 2PEs. As a result, the excitation spectrum shown in Fig.~\ref{spectra} for $a_{s}<0$ is only for 2PEs. In particular, it is found that the softened two-particle breathing mode in the CSC phase turns into the Goldstone zero mode of the SSC phase when $a_s$ becomes negative. Unlike that in the CSC phase, the lowest nonzero excitations in the SSC phase (labeled d) are two-particle dipole modes, which suggests that SSCs are less compressible than CSCs since any mode describing the deformation of a SSC costs more energy than the center-of-mass motion. We note that the deviation of the dipole excitation energy from $\hbar \omega _{\mathrm{ho}}$ for $a_{s}$ close to the stability boundary is because the number of the
basis states used in our numerical calculations is inadequate to ensure the
convergence of the solution~\cite{insufficient}. Finally, mode e in Fig.~\ref{spectra} is the two-particle breathing mode which softens with the increase of $|a_s|$ and eventually becomes unstable.

\textit{Discussion and conclusion}.---It is well-known that the CSC has
Poissonian statistics with particle number fluctuation $\Delta N=\sqrt{N}$.
In contrast, the particle number statistics of the SSC is super-Poissonian
which has a larger number fluctuation $\Delta N\approx \sqrt{2}N$.
Consequently, the SSC is also featured by the normalized second-order
correlation function
\end{subequations}
\begin{equation*}
g^{(2)}(0)=\frac{\big\langle\hat{\psi}^{\dagger 2}({\mathbf r})\hat{\psi}^{2}({\mathbf r})
\big\rangle}{\big|\big\langle\hat{\psi}^{\dagger }({\mathbf r})\hat{\psi}({\mathbf r})\big\rangle
\big|^{2}}\approx 3,
\end{equation*}
as compared to $g^{(2)}(0)\approx 1$ for CSCs. Because the interaction
energy density is proportional to $g^{(2)}(0)$, the large number fluctuation
of the SSCs leads to a lower interaction energy when $a_{s}<0$ and a tripled
interaction strength in the effective Eq.~\eqref{Ims}.

This unique features of the SSCs also offers clues for their experimental
detection. For instance, direct measurement of the particle-number statistics of an weakly attractive condensate would single a SSC out from a CSC. In addition, one may also measure the critical scattering length $N|a_{s}|/a_{\mathrm{ho}}$ which is around $0.19$ for SSCs. In fact,
we are aware of that a measured value in the $^{39}$K experiment
performed by LENS group was roughly $0.2$~\cite{K39}, which is very
likely due to the large particle number fluctuations in squeezed states. Additionally, the measurement of the Tan contact \cite{Tan,TanEx} across the gas-droplet phase transition may provide evidences for the change of
the interaction strength. Finally, a smoking-gun signature for the
squeezed-state condensate is provided by measuring $g^{(2)}(0)$ in
time-of-flight experiments. In fact, as shown in the Supplement Material~%
\cite{SM}, $g^{(2)}$ remain unchanged if the atom-atom interaction is
switched off in the free expansion.

In conclusion, we have shown that the quantum ground state of a weakly attractive
condensate is a single-mode squeezed state. An
effective equation governing the spatial mode of the SSCs has
also be derived, in which, due to the large number fluctuation of the
squeezed state, the interaction strength is tripled compared to that in the
conventional GPE. Our findings clarify a
widely accepted misconception about the quantum state of attractive
condensates. We believe that this study will open new avenues for research
in ultracold atomic gases, in particular, for the droplet phases in dipolar and
multiple-component condensates \cite{dipolarEx,twocomponentEx}. Our future
works will include the studies of droplet phases via GST. And, for
completeness, we shall also include three-body interactions in our studies.

We thank the enlightening discussion with J. Ignacio Cirac, Eugene Demler,
Carlos Navarrete-Benlloch, Xinyu Luo, Xingyan Chen, Li You, Chang-Pu Sun, and
Han Pu. TS acknowledges the Thousand-Youth-Talent Program of China. This work
was supported by the NSFC (Grants No. 11434011 and No. 11674334), by the
Strategic Priority Research Program of CAS (Grant No. XDB28000000), and by
National Key Research and Development Program of China (Grant No.
2017YFA0304501).

\onecolumngrid
\clearpage

\begin{center}
\textbf{\large Supplemental Material}
\end{center}

\setcounter{equation}{0} \setcounter{figure}{0} 
\makeatletter

\renewcommand{\thefigure}{SM\arabic{figure}} \renewcommand{\thesection}{SM%
\arabic{section}} \renewcommand{\theequation}{SM\arabic{equation}}

In Sec. SM1, we derive the ground state energy and mean-field
Hamiltonian using the Wick's theorem. Then we show in Sec. SM2 how to projection EOM onto the basis of a spherical harmonic oscillator, which can be used to solve the EOM numerically. In Sec. SM3, we show how to obtain the Lee-Huang-Yang (LHY) corrected Gross-Pitaevskii equation from the Gaussian-state theory. In Sec. SM4, we show in detail how to linearize the real-time EOM which allows us to systematically perform the fluctuation analysis. Finally, we calculate the second order correlation function $g^{(2)}(0)$ of a free expanded squeezed-state condensate in Sec. SM5.

\section{Derivation of the EOM}

In the section, we derive the ground state energy and the mean-field
Hamiltonian for the general Gaussian states. To this end, we decompose the
field operator into the coherent part $\phi ({\mathbf r})$ and the fluctuation
operator $\delta \hat{\psi}({\mathbf r})$, i.e., $\hat{\psi}({\mathbf r})=\phi ({\mathbf r})+\delta \hat{%
\psi}({\mathbf r})$. The total Hamiltonian of the system can then be expressed as $%
H=\sum_{j=0}^{4}h_{j}$, where
\begin{equation}
h_{0}=\int d{\mathbf r}\phi ^{\ast }({\mathbf r})\left[{\mathcal L}+\frac{U}{2}\left\vert
\phi ({\mathbf r})\right\vert ^{2}\right]\phi ({\mathbf r})
\end{equation}%
is the constant term,
\begin{eqnarray}
h_{1}+h_{3} &=&\int d{\mathbf r}\left[\delta \hat{\psi}^{\dagger }({\mathbf r}){\mathcal L}\phi
({\mathbf r})+U\delta \hat{\psi}^{\dagger }({\mathbf r})\left\vert \phi ({\mathbf r})\right\vert
^{2}\phi ({\mathbf r})+U\delta \hat{\psi}^{\dagger 2}({\mathbf r})\delta \hat{\psi}({\mathbf r})\phi
({\mathbf r})+\mathrm{H.c.}\right]
\end{eqnarray}%
are the linear and cubic terms, and
\begin{eqnarray}
h_{2}+h_{4} &=&\int d{\mathbf r}\left\{\delta \hat{\psi}^{\dagger }({\mathbf r})\left[%
{\mathcal L}+2U\left\vert \phi ({\mathbf r})\right\vert ^{2}\right]\delta \hat{\psi}%
({\mathbf r})+\frac{U}{2}\left[\phi ^{2}({\mathbf r})\delta \hat{\psi}^{\dagger 2}({\mathbf r})+%
\mathrm{H.c.}\right]+\frac{U}{2}\delta \hat{\psi}^{\dagger 2}({\mathbf r})\delta
\hat{\psi}^{2}({\mathbf r})\right\} 
\end{eqnarray}
are the quadratic and quartic terms. It follows from Wick's theorem that
the mean-field Hamiltonian, in normal ordered form, is
\begin{align}
H_{\mathrm{MF}}&=E+H_{1}+H_{2}, \\
E &=h_{0}+\int d{\mathbf r}\left\{\lim_{{\mathbf r}^{\prime }\rightarrow
{\mathbf r}}{\mathcal L}({\mathbf r})\left\langle \delta \hat{\psi}^{\dagger }({\mathbf r}^{\prime })\delta
\hat{\psi}({\mathbf r})\right\rangle+2U\left[\left\vert \phi ({\mathbf r})\right\vert ^{2}+%
\frac{1}{2}\left\langle \delta \hat{\psi}^{\dagger }({\mathbf r})\delta \hat{\psi}%
({\mathbf r})\right\rangle\right] \left\langle \delta \hat{\psi}^{\dagger }({\mathbf r})\delta
\hat{\psi}({\mathbf r})\right\rangle \right.  \notag \\
&\left.\qquad\qquad\qquad\quad+\frac{U}{2}\left[\phi ^{2}({\mathbf r})\left\langle
\delta \hat{\psi}^{\dagger 2}({\mathbf r})\right\rangle+{\rm H.c.}\right]+\frac{U}{2}%
\left\langle \delta \hat{\psi}^{2}({\mathbf r})\right\rangle \left\langle
\delta \hat{\psi}^{\dagger 2}({\mathbf r})\right\rangle \right\}, \\
H_{1}&=\int d{\mathbf r}\left[\delta \hat{\psi}^{\dagger }({\mathbf r})\eta ({\mathbf r})+\mathrm{H.c.}%
\right], \\
H_{2}&=\frac{1}{2}\int d{\mathbf r}:\!\delta \hat{\Psi} ^{\dagger }({\mathbf r})\mathcal{H}%
\delta \hat{\Psi}({\mathbf r})\!:,\label{sm7}
\end{align}%
where $\eta$ and $\mathcal{H}$ are given in the main text. We note that, in Eq.~\eqref{sm7}, the matrix multiplication in coordinate space in the term $\frac{1}{2}:\delta \hat{\Psi}^{\dagger }\mathcal{H}\delta \hat{\Psi}:$ of Eq.~\eqref{hmeanf} is explicitly expressed as integration over ${\mathbf r}$ in Eq.~\eqref{sm7}. Similarly, the products in Eqs.~\eqref{GS}%
, \eqref{Imv}, \eqref{Rev}, and \eqref{fRev} of the main text should also be understood as the matrix multiplications in the coordinate and Nambu spaces. The EOM in imaginary time [Eqs.~(\ref{Imv}) in the main text]
are obtained by projecting
\begin{equation}
\partial _{\tau }\left\vert \Psi \right\rangle =-(H-\left\langle
H\right\rangle )\left\vert \Psi \right\rangle
\end{equation}%
onto the tangential space of the variational Gaussian manifold \cite{Shi}.
In similar way, we obtain the EOM in the real time, i.e., Eqs. (\ref{Rev})
in the main text. In the next section, we demonstrate how to solve Eqs. (\ref%
{Imv}) numerically by expanding them onto the eigenbasis of the harmonic oscillator.

\section{EOM in the eigenbasis of an isotropic harmonic oscillator}

Here we show how to project the imaginary-time EOM onto the eigenbasis of a
spherical harmonic oscillator, which further allows us to solve the EOM numerically. To this end, we note that the
eigenstates of a spherical harmonic oscillator are
\begin{equation}
\varphi _{nlm}({\mathbf r})=R_{nl}(r)Y_{lm}(\Omega)=\sqrt{\frac{2\Gamma (n+l+3/2)}{%
n!\Gamma (l+3/2)^{2}}}r^{l}e^{-\frac{1}{2}r^{2}}F(-n,l+3/2,r^{2})Y_{lm}(%
\Omega )
\end{equation}%
with eigenenergies $\varepsilon_{nl}=2n+l+3/2-\mu$, where $n\geq 0$, $%
l=n,n-2,n-4,\ldots,1\mbox{ or }0$, $m=-l,-l+1,\ldots, l$, $F(a,b,z)$ is the
confluent hypergeometric function of the first kind, $Y_{lm}(\Omega)$ are
spherical harmonics, and we have used $a_{\mathrm{ho}}$ as length unit. Now,
the coherent part and the covariance matrix elements can be expanded in the
basis set $\{\varphi_{nlm}\}$ as
\begin{eqnarray}
\phi (x) &=&\sum_{n}\varphi _{n00}(r)\beta _{n}, \\
G(x,x^{\prime }) &=&\sum_{nn^{\prime }}\sum_{lm}\varphi _{n^{\prime
}lm}^{\ast }(r^{\prime })\varphi _{nlm}(r)G_{nn^{\prime }}^{lm}, \\
F(x,x^{\prime }) &=&\sum_{nn^{\prime }}\sum_{lm}\varphi _{n^{\prime
}l-m}(r^{\prime })\varphi _{nlm}(r)F_{nn^{\prime }}^{lm},
\end{eqnarray}%
where $\beta_n$, $G_{nn^{\prime }}^{lm}$, and $F_{nn^{\prime}}^{lm}$ are
expansion coefficients. Due to the rotational symmetry, we have $%
G_{nn^{\prime }}^{lm}=G_{nn^{\prime}}^{l}$ and $F_{nn^{%
\prime}}^{lm}=(-1)^{m}F_{nn^{\prime }}^{l}$. Namely, $G_{nn^{\prime }}^{lm}$
is independent of $m$ and $F_{nn^{\prime}}^{lm}$ only depends on $m$ through
the sign $(-1)^m$. After substituting these expansions into the EOM (\ref%
{Imv}), we obtain
\begin{align}
\partial _{\tau }\left(
\begin{array}{c}
\beta _{n} \\
\beta _{n}^{\ast }%
\end{array}%
\right) & =-\sum_{n^{\prime }}\Gamma _{nn^{\prime }}^{00}%
\begin{pmatrix}
\eta _{n^{\prime }} \\
\eta _{n^{\prime }}^{\ast }%
\end{pmatrix}%
,  \label{Ime1} \\
\partial _{\tau }\Gamma _{nn^{\prime }}^{lm}& =\sigma ^{z}\mathcal{H}%
_{nn^{\prime }}^{lm}\sigma ^{z}-\sum_{ss^{\prime }}\Gamma _{ns}^{lm}\mathcal{%
H}_{ss^{\prime }}^{lm}\Gamma _{s^{\prime }n^{\prime }}^{lm},  \label{Ime}
\end{align}
where the covariance matrix is
\begin{equation}
\Gamma _{nn^{\prime }}^{lm}=2\left(
\begin{array}{cc}
G_{nn^{\prime }}^{l} & (-1)^{m}F_{nn^{\prime }}^{l} \\
(-1)^{m}F_{nn^{\prime }}^{l} & G_{nn^{\prime }}^{l}%
\end{array}
\right) +\mathbf{I}
\end{equation}%
and the linear driving vector is
\begin{equation}
\eta _{n}=\varepsilon _{n0}\beta _{n}+\sum_{n^{\prime }n_{1}n_{1}^{\prime
}}\sum_{l_{1}}M_{nn^{\prime }n_{1}n_{1}^{\prime }}^{0,l_{1}}(2l_{1}+1)\left[%
(\beta _{n_{1}^{\prime }}^{\ast }\beta _{n_{1}}\delta
_{l_{1}0}+2G_{n_{1}n_{1}^{\prime }}^{l_{1}})\beta _{n^{\prime
}}+F_{n_{1}n_{1}^{\prime }}^{l_{1}}\beta _{n^{\prime }}^{\ast }\right].
\end{equation}
Moreover, the interaction matrix elements are
\begin{align}
M_{nn^{\prime }n_{1}n_{1}^{\prime }}^{ll_{1}} &=a_{s}\int_{0}^{\infty
}drr^{2}R_{nl}(r)R_{n^{\prime }l}(r)R_{n_{1}l_{1}}(r)R_{n_{1}^{\prime
}l_{1}}(r)  \notag \\
&=\frac{a_{s}}{2^{l+l_{1}+1/2}}%
\mbox{$\sqrt{n!n^{\prime }!n_{1}!n_{1}^{\prime }!\Gamma \left(n+l+\frac{3}{2}\right)\Gamma \left(n^{\prime
}+l+\frac{3}{2}\right)\Gamma \left(n_{1}+l_{1}+\frac{3}{2}\right)\Gamma \left(n_{1}^{\prime }+l_{1}+\frac{3}{2}\right)}$}
\notag \\
&\quad\times\sum_{k_{1}k_{2}k_{3}k_{4}}\Gamma \left(\frac{3}{2}+l+l_{1}+{K}%
\right)\frac{1}{(n-k_{1})!}\frac{1}{(n^{\prime }-k_{2})!}\frac{1}{%
(n_{1}-k_{3})!}\frac{1}{(n_{1}^{\prime }-k_{4})!}  \notag \\
&\quad\qquad\qquad\times\frac{1}{\Gamma \left(l+\frac{3}{2}%
+k_{1}\right)\Gamma \left(l+\frac{3}{2}+k_{2}\right)\Gamma \left(l_{1}+\frac{%
3}{2}+k_{3}\right)\Gamma \left(l_{1}+\frac{3}{2}+k_{4}\right)}\frac{(-\frac{1%
}{2})^{K}}{k_{1}!k_{2}!k_{3}!k_{4}!},
\end{align}%
where $K=\sum_ik_i$ and the summation over $k_i$ runs over all integers that
validate the term being summed. Finally, the mean-field Hamiltonian $%
\mathcal{H}$ is block diagonalized as $\mathcal{\bigoplus }_{lm}\mathcal{H}%
^{lm}$ in the eigenbasis $\varphi _{nlm}(r)$, where%
\begin{equation}
\mathcal{H}^{lm}=\left(
\begin{array}{cc}
\mathcal{E}^{l} & (-1)^{m}\Delta ^{l} \\
(-1)^{m}\Delta ^{l\ast } & \mathcal{E}^{l\ast }%
\end{array}%
\right)
\end{equation}%
with
\begin{eqnarray}
\mathcal{E}^{l} &=&\varepsilon _{nl}\delta _{nn^{\prime
}}+2\sum_{n_{1}n_{1}^{\prime },l_1}M_{nn^{\prime }n_{1}n_{1}^{\prime
}}^{l,l_{1}}(2l_{1}+1)\left(\delta _{l_{1}0}\beta _{n_{1}^{\prime }}^{\ast }\beta
_{n_{1}}+G_{n_{1}n_{1}^{\prime }}^{l_{1}}\right), \\
\Delta ^{l} &=&\sum_{n_{1}n_{1}^{\prime },l_1}M_{nn^{\prime
}n_{1}n_{1}^{\prime }}^{l,l_{1}}(2l_{1}+1)\left(\delta _{l_{1}0}\beta
_{n_{1}^{\prime }}\beta _{n_{1}}+F_{n_{1}n_{1}^{\prime }}^{l_{1}}\right).
\label{HL}
\end{eqnarray}

In numerical calculation, the basis set $\{\varphi_{nlm}\}$ is truncated by
introducing $n_{\mathrm{cut}}$ and letting $n\leq n_{\mathrm{cut}}$. The
ground state solution can then be obtain by numerically evolving Eqs. (\ref%
{Ime1}) and (\ref{Ime}) until $\beta_n$, $G_{nn^{\prime }}^{l}$, and $%
F_{nn^{\prime}}^{l}$ all converge.

\section{Lee-Huang-Yang correction derived from Gaussian-state theory}

Here we show in detail how to derive the LHY correction from the Gaussian-state theory. For positive scattering length, the optimal Gaussian state is close to a coherent state. By neglecting the contributions of $\left\langle \delta \hat{%
\psi}^{\dagger }(x)\delta \hat{\psi}(x)\right\rangle $ and $\left\langle
\delta \hat{\psi}^{2}(x)\right\rangle $, we find
\begin{align}
\eta &=\left[{\mathcal L}+U\left\vert \phi ({\mathbf r})\right\vert ^{2}\right]\phi({\mathbf r}), \\
\mathcal{E}&={\mathcal L}+2U\left\vert \phi ({\mathbf r})\right\vert ^{2}, \\
\Delta&=U\phi ^{2}({\mathbf r}).
\end{align}%
The steady-state condition, $\partial _{\tau }\Phi =0$, then leads to the GP
equation%
\begin{equation}
\left[{\mathcal L}+U\left\vert \phi({\mathbf r})\right\vert ^{2}\right]\phi({\mathbf r})=0,
\label{GP}
\end{equation}
whose solution gives rise to the ground state wavefunction $\phi_{0}({\mathbf r})$ of
the coherent-state condensate. Another steady-state condition $\partial
_{\tau }\Gamma =0$, namely,
\begin{equation}
\Sigma ^{z}\mathcal{H}[\phi _{0},\Gamma _{0}]\Sigma ^{z}-\Gamma _{0}\mathcal{%
H}[\phi _{0},\Gamma _{0}]\Gamma _{0}=0,  \label{Gamma}
\end{equation}%
can be used to determine $\Gamma_0$ which describes the small depletion and
squeezing.

To proceed, let us first show that Eq. (\ref{Gamma}) stands if $\mathcal{H}%
[\phi _{0},\Gamma _{0}]$ can be diagonalized by a symplectic matrix $S_0$
(defined by $S_{0}\Sigma ^{z}S_{0}^{\dagger }=\Sigma ^{z}$), i.e., $%
S_{0}^{\dagger }\mathcal{H}[\phi _{0},\Gamma_{0}]S_{0}=I_{2}\otimes D$,
where $\Gamma_0$ is fixed by $\Gamma_0=S_0S_0^\dagger$. The proof goes as
follows.
\begin{align}
\Sigma ^{z}\mathcal{H}[\phi _{0},\Gamma _{0}]\Sigma ^{z}-\Gamma _{0}
\mathcal{H}[\phi _{0},\Gamma _{0}]\Gamma _{0}&=\Sigma ^{z}\mathcal{H}[\phi
_{0},\Gamma _{0}]\Sigma ^{z}-S_{0}S_{0}^{\dagger }\mathcal{H}[\phi
_{0},\Gamma _{0}]S_{0}S_{0}^{\dagger }  \notag \\
&=\Sigma ^{z}\mathcal{H}[\phi _{0},\Gamma _{0}]\Sigma
^{z}-S_{0}(I_{2}\otimes D)S_{0}^{\dagger }  \notag \\
&=\Sigma ^{z}\mathcal{H}[\phi _{0},\Gamma _{0}]\Sigma ^{z}-\Sigma ^{z}%
\mathcal{H}[\phi _{0},\Gamma _{0}]\Sigma ^{z}  \notag \\
&=0.
\end{align}

For the homogeneous system, Eq. (\ref{GP}) gives the relation $\mu
=Un_{0}$ of the chemical potential and the condensate density $%
n_{0}=\left\vert \phi _{0}\right\vert ^{2}$. The Bogoliubov transformation%
\begin{equation}
S_{k}=\left(
\begin{array}{cc}
u_{k} & v_{k} \\
v_{k} & u_{k}%
\end{array}%
\right)
\end{equation}%
diagonalizes the mean-field Hamiltonian%
\begin{equation}
\mathcal{E}_{k}=\frac{{\mathbf k}^{2}}{2m}+n_{0}U^{(d)},\quad \Delta _{k}=n_{0}U^{(d)}
\end{equation}%
in the momentum space, where the Bogoliubov parameters%
\begin{equation}
u_{k}=\sqrt{\frac{1}{2}\left(\frac{\mathcal{E}_{k}}{E_{k}}+1\right)},\quad
v_{k}=-\sqrt{\frac{1}{2}\left(\frac{\mathcal{E}_{k}}{E_{k}}-1\right)}
\end{equation}%
are determined by the single excitation spectrum $E_{k}=\sqrt{\mathcal{E}%
_{k}^{2}-\Delta _{k}^{2}}$. The depletion and the squeezing effects can be evaluated as 
\begin{eqnarray}
\left\langle \delta \hat{\psi}^{\dagger }({\mathbf r})\delta \hat{\psi}%
({\mathbf r})\right\rangle &=&\int \frac{d{\mathbf k}}{(2\pi )^{3}}\frac{1}{2}\left(\frac{%
\mathcal{E}_{k}}{E_{k}}-1\right)=\frac{8}{3}\sqrt{\frac{n_{0}^{3}a_{s}^{3}}{%
\pi }},  \notag \\
\left\langle \delta \hat{\psi}^{2}({\mathbf r})\right\rangle &=&-\int \frac{d{\mathbf k}}{%
(2\pi )^{3}}\Delta _{k}\left(\frac{1}{2E_{k}}-\frac{m}{k^{2}}\right)=8\sqrt{%
\frac{n_{0}^{3}a_{s}^{3}}{\pi }}.
\end{eqnarray}%
In the first order iteration, we take into account the contributions of $%
\left\langle \delta \hat{\psi}^{\dagger }({\mathbf r})\delta \hat{\psi}%
({\mathbf r})\right\rangle $ and $\left\langle \delta \hat{\psi}^{2}({\mathbf r})\right\rangle $
in $\eta $, which leads to the chemical potential $\mu =n_{0}U+\delta\mu $, where
\begin{equation}
\delta \mu =\frac{40}{3}n_{0}U^{(3)}\sqrt{\frac{n_{0}a_{s}^{3}}{\pi }}
\end{equation}
is the LHY correction to the chemical potential. For the inhomogeneous system with slowly varying potentials, the
local density approximation then leads to the GP equation with LHY corrections
\begin{equation}
\left[{\mathcal L}+U\left\vert \phi _{0}({\mathbf r})\right\vert ^{2}+\frac{40}{3}%
Ua_{s}\sqrt{\frac{a_{s}}{\pi }}\left\vert \phi _{0}({\mathbf r})\right\vert ^{3}%
\right]\phi _{0}({\mathbf r})=0.
\end{equation}

\section{Fluctuation analysis}

To obtain the fluctuation spectrum, we linearize EOM (\ref{Rev}) around a
steady state solution $\phi _{0}({\mathbf r})$ and $\Gamma _{0}=S_{0}S_{0}^{\dagger }$
of Eq. (\ref{Imv}). Specifically, we decompose the coherent part and
two-particle excitations described by $\delta \phi ({\mathbf r})$ and $\delta
\vartheta $, respectively. In the second quantized form, the state with
fluctuation reads%
\begin{equation}
\left\vert \bar{\Psi}_{\mathrm{GS}}\right\rangle =e^{\hat{\Psi}^{\dagger
}\Sigma ^{z}(\Phi +\delta \Phi )}e^{i\frac{1}{2}\hat{\Psi}^{\dagger }\xi _{0}%
\hat{\Psi}}e^{i\frac{1}{2}\hat{\Psi}^{\dagger }\delta \vartheta \hat{\Psi}%
}\left\vert 0\right\rangle ,
\end{equation}%
where $\delta \Phi =(\delta \phi ,\delta \phi ^{\ast })^T$. Correspondingly,
the symplectic matrix becomes $S=S_{0}e^{i\Sigma ^{z}\delta \vartheta }$.
However, there is some gauge redundancy \cite{Tommaso} in the generator $%
\delta \vartheta $ that does not change the covariance matrix, or
equivalently, the ground state. This can be seen as follows. Considering the
infinitesimal generator $\delta \vartheta $, we can expnd $S\sim
S_{0}(1+i\Sigma ^{z}\delta \vartheta )$ to the first order. Since $\Gamma
=SS^{\dagger }\sim \Gamma _{0}+iS_{0}[\Sigma ^{z},\delta \vartheta
]S_{0}^{\dagger }$, the generator that commutes with $\Sigma ^{z}$ does not
change the covariance matrix. In general, the generator $\delta \vartheta $
without redundancy only has the off-diagonal form $\delta \vartheta =\left(
\begin{array}{cc}
0 & \delta \xi \\
\delta \xi ^{\dagger } & 0%
\end{array}%
\right) $. The symplectic matrix the becomes
\begin{equation}
S=S_{0}\exp \left[ i\Sigma ^{z}\left(
\begin{array}{cc}
0 & \delta \xi \\
\delta \xi ^{\dagger } & 0%
\end{array}%
\right) \right] ,
\end{equation}%
and the fluctuation of the covariance matrix becomes%
\begin{equation}
\delta \Gamma =\Gamma -\Gamma _{0}=SS^{\dagger }-S_{0}S_{0}^{\dagger
}=2iS_{0}\left(
\begin{array}{cc}
0 & \delta \xi \\
-\delta \xi ^{\dagger } & 0%
\end{array}%
\right) S_{0}^{\dagger }.  \label{dG}
\end{equation}

In terms of $S$, the EOM (\ref{Rev}) can be written in the equivalent form
\begin{eqnarray}
i\partial _{t}\phi &=&\eta ,  \label{RevS1} \\
iS^{\dagger }\Sigma ^{z}\partial _{t}S &=&S^{\dagger }\mathcal{H}S.
\label{RevS}
\end{eqnarray}%
The linearization of Eq. (\ref{RevS1}) and (\ref{RevS}) around $\phi _{0}$
and $S_{0}$ results in
\begin{eqnarray}
i\partial _{t}\delta \phi &=&\delta \eta ,  \label{fluc1} \\
i\partial _{t}\delta \xi &=&\{D,\delta \xi \}-i(S_{0}^{\dagger }\delta
\mathcal{H}S_{0})_{12},  \label{fluc}
\end{eqnarray}%
where the subscript $12$ denotes the off-diagonal block in the Nambu basis,
and the fluctuations of the linear driving term and the mean-field
Hamiltonian are%
\begin{equation}
\delta \eta =\mathcal{E}\delta \phi ({\mathbf r})+\Delta \delta \phi ^{\ast
}({\mathbf r})+2U\phi _{0}({\mathbf r})\delta G({\mathbf r},{\mathbf r})+U\phi _{0}^{\ast }({\mathbf r})\delta
F({\mathbf r},{\mathbf r}),
\end{equation}%
and $\delta \mathcal{H}=\left(
\begin{array}{cc}
\delta \mathcal{E} & \delta \Delta \\
\delta \Delta ^{\dagger } & \delta \mathcal{E}^{\ast }%
\end{array}%
\right) $:%
\begin{eqnarray}
\delta \mathcal{E} &=&2U[\phi _{0}^{\ast }({\mathbf r})\delta \phi ({\mathbf r})+\phi
_{0}({\mathbf r})\delta \phi ^{\ast }({\mathbf r})+\delta G({\mathbf r},{\mathbf r})], \\
\delta \Delta &=&U[2\phi _{0}({\mathbf r})\delta \phi ({\mathbf r})+\delta F({\mathbf r},{\mathbf r})].
\end{eqnarray}

As an example, we demonstrate how to diagonalize Eqs.~\eqref{fluc1} and %
\eqref{fluc} numerically in a 3D spherical trap. Assuming that a
steady-state solution $\mathcal{(}\phi _{0},\Gamma _{0})$ is numerically
obtained via the method introducing in SM2, which allows us to construct the
mean-field Hamiltonian $\mathcal{H}[\phi _{0},\Gamma _{0}]$. The symplectic
diagonlization of $\mathcal{H}[\phi _{0},\Gamma _{0}]$ then gives rise to
the spectrum $D$ and $S_{0}$, where in the eigenbasis $\phi
_{0}({\mathbf r})=\sum_{n}\varphi _{n00}(r)\beta _{n}^{(0)}$ and%
\begin{equation}
S_{0}=\sum_{lm}\varphi _{nlm}(r)\left(
\begin{array}{cc}
u_{ns}^{l} & (-1)^{m}v_{ns}^{l\ast } \\
v_{ns}^{l} & (-1)^{m}u_{ns}^{l\ast }%
\end{array}%
\right)  \label{S0}
\end{equation}%
is determined by the Bogoliubov parameters $u_{ns}^{l}$ and $v_{ns}^{l}$.

Since the total angular momentum is conserved, we consider the
single-Bogoliubov excitation $\delta \phi ({\mathbf r})=\varphi _{nLM_{L}}(r)\delta
\beta _{nLM_{L}}$ with angular momemntum ($L,M_{L}$). It is clear that in
the second quantized form, $\delta \xi _{s_{1},s_{1}^{\prime
}}^{l_{1}m_{1},l_{1}^{\prime }m_{1}^{\prime }}$ is the wavefunction of the
two-excitation state $\sum_{s_{1}s_{1}^{\prime },l_{1}m_{1}l_{1}^{\prime
}m_{1}^{\prime }}\delta \xi _{s_{1},s_{1}^{\prime
}}^{l_{1}m_{1},l_{1}^{\prime }m_{1}^{\prime }}b_{s_{1}l_{1}m_{1}}^{\dagger
}b_{s_{1}^{\prime }l_{1}^{\prime }-m_{1}^{\prime }}^{\dagger }\left\vert
0\right\rangle /2$, where $b_{s_{1}l_{1}m_{1}}^{\dagger }b_{s_{1}^{\prime
}l_{1}^{\prime }-m_{1}^{\prime }}^{\dagger }$ creates two Bogoliubov
excitaitons with angular momenta $l_{1}m_{1}$ and $l_{1}^{\prime
}m_{1}^{\prime }$. Thus, two Bogoliubov excitations with the total angular
momentum ($L,M_{L}$) are described by $\delta \xi _{s_{1},s_{1}^{\prime
}}^{l_{1}m_{1},l_{1}^{\prime }m_{1}^{\prime }}=\sqrt{2}C_{l_{1}m_{1},l_{1}^{%
\prime }-m_{1}^{\prime }}^{LM_{L}}\delta \xi _{s_{1},s_{1}^{\prime
}}^{l_{1},l_{1}^{\prime }}$, where $C_{l_{1}m_{1},l_{1}^{\prime
}m_{1}^{\prime }}^{LM_{L}}$ is the Clebsch-Gordan coefficients.

Due to the rotational symmetry along the $z$ axis, the spectrum of the
excitation with momentum $L$ has $(2L+1)$-fold degeneracy, thus without loss
of generality we can choose $M_{L}=0$. The fluctuation%
\begin{align}
\delta G({\mathbf r},{\mathbf r})& =i\sum_{m_{1}}\sum_{l_{1}l_{1}^{\prime
}}\sum_{n_{1}n_{1}^{\prime }s_{1},s_{1}^{\prime }}\varphi
_{n_{1}l_{1}m_{1}}(r)\varphi _{n_{1}^{\prime }l_{1}^{\prime
}-m_{1}}(r)\left( u_{n_{1}s_{1}}^{l_{1}}v_{n_{1}^{\prime }s_{1}^{\prime
}}^{l_{1}^{\prime }}\delta \xi _{s_{1},s_{1}^{\prime
}}^{l_{1}m_{1},l_{1}^{\prime }m_{1}}-v_{n_{1}s_{1}}^{l_{1}\ast
}u_{n_{1}^{\prime }s_{1}^{\prime }}^{l_{1}^{\prime }\ast }\delta \xi
_{s_{1},s_{1}^{\prime }}^{l_{1}m_{1},l_{1}^{\prime }m_{1}\ast }\right) , \\
\delta F({\mathbf r},{\mathbf r})& =i\sum_{m_{1}}\sum_{l_{1}l_{1}^{\prime
}}\sum_{n_{1}n_{1}^{\prime }s_{1},s_{1}^{\prime }}\varphi
_{n_{1}l_{1}m_{1}}(r)\varphi _{n_{1}^{\prime }l_{1}^{\prime
}-m_{1}}(r)\left( u_{n_{1}s_{1}}^{l_{1}}u_{n_{1}^{\prime }s_{1}^{\prime
}}^{l_{1}^{\prime }}\delta \xi _{s_{1},s_{1}^{\prime
}}^{l_{1}m_{1},l_{1}^{\prime }m_{1}}-v_{n_{1}s_{1}}^{l_{1}\ast
}v_{n_{1}^{\prime }s_{1}^{\prime }}^{l_{1}^{\prime }\ast }\delta \xi
_{s_{1},s_{1}^{\prime }}^{l_{1}m_{1},l_{1}^{\prime }m_{1}\ast }\right) ,
\end{align}%
of the correlation functions, i.e., the elemnets of $\delta \Gamma $, are
determined by Eqs. (\ref{dG}) and (\ref{S0}).

It follows from Eq. (\ref{fluc1}) that%
\begin{equation}
i\partial _{t}\delta \beta _{nL0}=\delta \eta _{nL},  \label{db}
\end{equation}%
where%
\begin{align}
\delta \eta _{nL} &=\mathcal{E}^{L}\delta \beta _{nL0}+\Delta ^{L}\delta
\beta _{nL0}^{\ast }  \notag \\
&\quad+i\sqrt{2}\sum_{l_{1}l_{1}^{\prime }s_{1},s_{1}^{\prime }}\sum_{n^{\prime
}n_{1}n_{1}^{\prime }}\beta _{n^{\prime }}^{(0)}\bar{M}_{nn^{\prime
}n_{1}n_{1}^{\prime }}^{L0l_{1}l_{1}^{\prime
}}\left[\left(u_{n_{1}s_{1}}^{l_{1}}v_{n_{1}^{\prime }s_{1}^{\prime }}^{l_{1}^{\prime
}}+v_{n_{1}s_{1}}^{l_{1}}u_{n_{1}^{\prime }s_{1}^{\prime }}^{l_{1}^{\prime
}}+u_{n_{1}s_{1}}^{l_{1}}u_{n_{1}^{\prime }s_{1}^{\prime }}^{l_{1}^{\prime
}}\right)\delta \xi _{s_{1},s_{1}^{\prime }}^{l_{1},l_{1}^{\prime }}\right.  \notag \\
&\left.\quad-\left(u_{n_{1}s_{1}}^{l_{1}\ast }v_{n_{1}^{\prime }s_{1}^{\prime
}}^{l_{1}^{\prime }\ast }+v_{n_{1}s_{1}}^{l_{1}\ast }u_{n_{1}^{\prime
}s_{1}^{\prime }}^{l_{1}^{\prime }\ast }+v_{n_{1}s_{1}}^{l_{1}\ast
}v_{n_{1}^{\prime }s_{1}^{\prime }}^{l_{1}^{\prime }\ast }\right)\delta \xi
_{s_{1},s_{1}^{\prime }}^{l_{1},l_{1}^{\prime }\ast }\right]
\end{align}%
is determined by ($\mathcal{E}^{L}$, $\Delta ^{L}$) [see Eq. (\ref{HL})] and%
\begin{equation}
\bar{M}_{nn^{\prime }n_{1}n_{1}^{\prime }}^{ll^{\prime }l_{1}l_{1}^{\prime
}}=C_{l0l^{\prime }0}^{L0}\sqrt{\frac{(2l+1)(2l^{\prime }+1)}{(2L+1)}}%
M_{nn^{\prime }n_{1}n_{1}^{\prime }}^{ll^{\prime }l_{1}l_{1}^{\prime }}\sqrt{%
\frac{(2l_{1}+1)(2l_{1}^{\prime }+1)}{(2L+1)}}C_{l_{1}0l_{1}^{\prime }0}^{L0}
\end{equation}%
is determined by%
\begin{eqnarray}
M_{nn^{\prime }n_{1}n_{1}^{\prime }}^{ll^{\prime }l_{1}l_{1}^{\prime }} &=&%
\frac{a_{s}}{2^{(l+l^{\prime }+l_{1}+l_{1}^{\prime }+1)/2}}\sqrt{n!n^{\prime
}!n_{1}!n_{1}^{\prime }!\Gamma (n+l+3/2)\Gamma (n^{\prime }+l^{\prime
}+3/2)\Gamma (n_{1}+l_{1}+3/2)\Gamma (n_{1}^{\prime }+l_{1}^{\prime }+3/2)}
\notag \\
&&\sum_{k_{1}k_{2}k_{3}k_{4}}\Gamma (\frac{3+l+l^{\prime
}+l_{1}+l_{1}^{\prime }}{2}+K)\frac{1}{(n-k_{1})!}\frac{1}{(n^{\prime
}-k_{4})!}\frac{1}{(n_{1}-k_{3})!}\frac{1}{(n_{1}^{\prime }-k_{2})!}  \notag
\\
&&\frac{1}{\Gamma (l+3/2+k_{1})\Gamma (l^{\prime }+3/2+k_{4})\Gamma
(l_{1}+3/2+k_{3})\Gamma (l_{1}^{\prime }+3/2+k_{2})}\frac{(-\frac{1}{2})^{K}%
}{k_{1}!k_{2}!k_{3}!k_{4}!}.
\end{eqnarray}%
Here, we have used the relation%
\begin{equation}
\sum_{m_{1}}C_{l_{1}m_{1},l_{1}^{\prime
}-m_{1}}^{L0}Y_{l_{1}m_{1}}(\hat {\mathbf r})Y_{l_{1}^{\prime }-m_{1}}(\hat {\mathbf r})=\sqrt{\frac{%
(2l_{1}+1)(2l_{1}^{\prime }+1)}{4\pi (2L+1)}}C_{l_{1}0l_{1}^{\prime
}0}^{L0}Y_{L0}(\hat {\mathbf r}).
\end{equation}%
The EOM (\ref{fluc}) gives rise to%
\begin{align}
i\partial _{t}\delta \xi _{s,s^{\prime }}^{l,l^{\prime }}
&=(D_{l}+D_{l^{\prime }})\delta \xi _{s,s^{\prime }}^{l,l^{\prime
}}\nonumber\\
&\quad+\sum_{l_{1}l_{1}^{\prime }s_{1}s_{1}^{\prime }}\sum_{nn^{\prime
}n_{1}n_{1}^{\prime }}\bar{M}_{nn^{\prime }n_{1}n_{1}^{\prime }}^{ll^{\prime
}l_{1}l_{1}^{\prime }}\left[\left(u_{sn}^{l\dagger }v_{s^{\prime }n^{\prime
}}^{l^{\prime }\dagger }+v_{sn}^{l\dagger }u_{s^{\prime }n^{\prime
}}^{l^{\prime }\dagger }\right)\left(u_{n_{1}s_{1}}^{l_{1}}v_{n_{1}^{\prime
}s_{1}^{\prime }}^{l_{1}^{\prime }}+v_{n_{1}s_{1}}^{l_{1}}u_{n_{1}^{\prime
}s_{1}^{\prime }}^{l_{1}^{\prime }}\right)\right.  \notag \\
&\left.\qquad\qquad\qquad\qquad\qquad\qquad\quad+u_{sn}^{l\dagger }u_{s^{\prime }n^{\prime }}^{l^{\prime }\dagger
}u_{n_{1}s_{1}}^{l_{1}}u_{n_{1}^{\prime }s_{1}^{\prime }}^{l_{1}^{\prime
}}+v_{sn}^{l\dagger }v_{s^{\prime }n^{\prime }}^{l^{\prime }\dagger
}v_{n_{1}s_{1}}^{l_{1}}v_{n_{1}^{\prime }s_{1}^{\prime }}^{l_{1}^{\prime
}}\right]\delta \xi _{s_{1},s_{1}^{\prime }}^{l_{1},l_{1}^{\prime }}  \notag \\
&\quad-\sum_{l_{1}l_{1}^{\prime }s_{1}s_{1}^{\prime }}\sum_{nn^{\prime
}n_{1}n_{1}^{\prime }}\bar{M}_{nn^{\prime }n_{1}n_{1}^{\prime }}^{ll^{\prime
}l_{1}l_{1}^{\prime }}\left[\left(u_{sn}^{l\dagger }v_{s^{\prime }n^{\prime
}}^{l^{\prime }\dagger }+v_{sn}^{l\dagger }u_{s^{\prime }n^{\prime
}}^{l^{\prime }\dagger }\right)\left(u_{n_{1}s_{1}}^{l_{1}\ast }v_{n_{1}^{\prime
}s_{1}^{\prime }}^{l_{1}^{\prime }\ast }+v_{n_{1}s_{1}}^{l_{1}\ast
}u_{n_{1}^{\prime }s_{1}^{\prime }}^{l_{1}^{\prime }\ast }\right)\right.  \notag \\
&\left.\qquad\qquad\qquad\qquad\qquad\qquad\quad+u_{sn}^{l\dagger }u_{s^{\prime }n^{\prime }}^{l^{\prime }\dagger
}v_{n_{1}s_{1}}^{l_{1}\ast }v_{n_{1}^{\prime }s_{1}^{\prime
}}^{l_{1}^{\prime }\ast }+v_{sn}^{l\dagger }v_{s^{\prime }n^{\prime
}}^{l^{\prime }\dagger }u_{n_{1}s_{1}}^{l_{1}\ast }u_{n_{1}^{\prime
}s_{1}^{\prime }}^{l_{1}^{\prime }\ast }\right]\delta \xi _{s_{1},s_{1}^{\prime
}}^{l_{1},l_{1}^{\prime }\ast }  \notag \\
&\quad-i\sqrt{2}\sum_{nn^{\prime }n_{1}n_{1}^{\prime }}\left(u_{sn}^{l\dagger
}v_{s^{\prime }n^{\prime }}^{l^{\prime }\dagger }+v_{sn}^{l\dagger
}u_{s^{\prime }n^{\prime }}^{l^{\prime }\dagger }+u_{sn}^{l\dagger
}u_{s^{\prime }n^{\prime }}^{l^{\prime }\dagger }\right)\bar{M}_{nn^{\prime
}n_{1}n_{1}^{\prime }}^{ll^{\prime }L0}\beta _{n_{1}^{\prime }}^{(0)}\delta
\beta _{n_{1}L0}  \notag \\
&\quad-i\sqrt{2}\sum_{nn^{\prime }n_{1}n_{1}^{\prime }}\left(u_{sn}^{l\dagger
}v_{s^{\prime }n^{\prime }}^{l^{\prime }\dagger }+v_{sn}^{l\dagger
}u_{s^{\prime }n^{\prime }}^{l^{\prime }\dagger }+v_{sn}^{l\dagger
}v_{s^{\prime }n^{\prime }}^{l^{\prime }\dagger }\right)\bar{M}_{nn^{\prime
}n_{1}n_{1}^{\prime }}^{ll^{\prime }L0}\beta _{n_{1}^{\prime }}^{(0)}\delta
\beta _{n_{1}L0}^{\ast }  \label{dxi}
\end{align}%
for the wavefucntion $\delta \xi _{s_{1},s_{1}^{\prime
}}^{l_{1},l_{1}^{\prime }}$ of two-Bogoliubov excitations.

The linearized Eqs. (\ref{db}) and (\ref{dxi}) show that $\delta \beta
_{nL0} $ and $\delta \xi _{s,s^{\prime }}^{l,l^{\prime }}$ couple to their
conjugate amplitudes, which together with their conjugate counterpart form
the generalized Bogoliubov fluctuation theory \cite{Tommaso}. In the gas
phase, the non-zero condensate part $\beta _{n}^{(0)}\neq 0$ induces the
coupling between the single- and two- Bogoliubov excitations, which
reproduces the Goldstone zero mode and the decay of the single excitation.
In the droplet phase, $\beta _{n}^{(0)}=0$, and the two excitation decouples
with the single Bogoliubov excitation.

\section{Second-order correlation functions in free expansion}

Here we evaluate the second-order correlation functions of a squeezed-state
condensate in the time-of-flight experiment. To this end, we tune the
scattering length to zero and switch off the trapping potential at time $t=0$%
. The system then executes free expansion governed by the Hamiltonian
\begin{equation}
H_{\mathrm{free}}=-\frac{1}{2m}\int d{\mathbf r}\psi ^{\dagger }({\mathbf r})\nabla^{2}\psi ({\mathbf r}).
\end{equation}%
At time $t=T$, the second-order correlation function becomes
\begin{equation}
g^{(2)}(0)=\frac{\left\langle \hat{\psi}^{\dagger 2}({\mathbf r})\hat{\psi}%
^{2}({\mathbf r})\right\rangle _{T}}{\left\vert \left\langle \hat{\psi}^{\dagger }({\mathbf r})%
\hat{\psi}({\mathbf r})\right\rangle _{T}\right\vert ^{2}},
\end{equation}%
where the expectation value is taken with respect to the state $\left\vert
\Psi (T)\right\rangle =e^{-iH_{\mathrm{free}}T}\left\vert \Psi _{\mathrm{ss}%
}\right\rangle $. Recalling that $\left\vert \Psi _{\mathrm{ss}%
}\right\rangle =e^{\frac{1}{2}\xi _{0}(b^{\dagger 2}-b^{2})}\left\vert
0\right\rangle $ with $b^{\dagger }=\int d{\mathbf r}f({\mathbf r})\hat{\psi}^{\dagger }({\mathbf r})$, the wave function at time $T$ can be obtained analytically as%
\begin{equation}
\left\vert \Psi (T)\right\rangle =e^{-iH_{\mathrm{free}}T}\left\vert \Psi _{\mathrm{ss}}\right\rangle =e^{\frac{1}{2}\xi _{0}(b_{T}^{\dagger
2}-b_{T}^{2})}\left\vert 0\right\rangle ,
\end{equation}%
where $b_{T}^{\dagger }=\int d{\mathbf r}f({\mathbf r},T)\hat{\psi}^{\dagger }({\mathbf r})$ with $%
f({\mathbf r},T)$ being the mode function that is determined by
\begin{equation}
f({\mathbf r},T)=\int d{\mathbf r}^{\prime }G({\mathbf r}-{\mathbf r}^{\prime },T)f({\mathbf r}^{\prime }),
\end{equation}%
Here,
\begin{equation}
G({\mathbf r}-{\mathbf r}^{\prime })=\int \frac{d{\mathbf k}}{(2\pi )^3}e^{-i\frac{{\mathbf k}^{2}}{2m}%
T+i{\mathbf k}\cdot({\mathbf r}-{\mathbf r}^{\prime })}=\left(\frac{m}{2\pi iT}\right)^{d/2}e^{i\frac{m}{2T}%
\left\vert {\mathbf r}-{\mathbf r}^{\prime }\right\vert ^{2}}
\end{equation}%
is the free-space propagator.

Making use of the fact that $\left\vert \Psi (T)\right\rangle $ is still a
single-mode squeezed state, one can easily evaluate the the rank-$1$
correlation functions
\begin{eqnarray}
\left\langle \hat{\psi}^{\dagger }({\mathbf r})\hat{\psi}({\mathbf r})\right\rangle
_{T}=N\left\vert f({\mathbf r},T)\right\vert ^{2}\mbox{ and } \left\langle \hat{\psi}%
^{2}({\mathbf r})\right\rangle _{T} =\sqrt{N(N+1)}f^{2}({\mathbf r},T),
\end{eqnarray}%
which lead to
\begin{eqnarray}
g^{(2)}(0) &=&\frac{2\left\langle \hat{\psi}^{\dagger }({\mathbf r})\hat{\psi}%
({\mathbf r})\right\rangle _{T}^{2}+\left\vert \left\langle \hat{\psi}%
^{2}({\mathbf r})\right\rangle _{T}\right\vert ^{2}}{\left\langle \hat{\psi}^{\dagger
}({\mathbf r})\hat{\psi}({\mathbf r})\right\rangle _{T}^{2}}  \notag \\
&=&3+\frac{1}{N}\sim 3
\end{eqnarray}%
by the Wick's theorem.


\begin{thebibliography}{99}
\bibitem{insufficient} Close to the stability boundary of the negative
scattering length, a slight increase of the interaction strength requires a
significant increment the number of the basis states to ensure the
convergence of the solution, which is beyond computer resource available to
us.
\end{thebibliography}
\end{document}